\documentclass[%
 reprint,
 superscriptaddress,
 amsmath,amssymb,
 aps,longbibliography
]{revtex4-1}

\usepackage{graphicx}
\usepackage{dcolumn}
\usepackage{bm}
\usepackage{hyperref}
\usepackage[mathlines]{lineno}
\usepackage[bottom]{footmisc}
\usepackage[super]{nth}%

\begin{document}
\raggedbottom


\title{Light Localization and Cooperative Coupling Effects in Aperiodic Vogel Spirals}

\author{F. Sgrignuoli}
\affiliation{Department of Electrical and Computer Engineering, Boston University, Boston, Massachusetts, 02215, USA.}
\author{R.Wang}
\affiliation{Department of Electrical and Computer Engineering, Boston University, Boston, Massachusetts, 02215, USA.}
\author{F. A. Pinheiro}
\affiliation{Instituto de Fisica, Universidade Federal do Rio de Janeiro, Rio de Janeiro-RJ 21941-972, Brazil }
\author{L. Dal Negro}
\email{dalnegro@bu.edu}
\affiliation{Department of Electrical and Computer Engineering, Boston University, Boston, Massachusetts, 02215, USA.}
\affiliation{Division of Material Science and Engineering, Boston University, Boston, Massachusetts, 02215, USA.}
\affiliation{Department of Physics, Boston University, Boston, Massachusetts, 02215, USA}
\begin{abstract}
By using the dyadic Green's matrix spectral method, we demonstrate that aperiodic deterministic Vogel spirals made of electric dipoles support light localization in three dimensions, an effect that does not occur in traditional uniform random media. We discover a light localization transition in  Vogel spiral arrays embedded in three-dimensional space by evaluating the Thouless conductance, the level spacing statistics, and by performing a finite-size scaling. This light localization transition is different from the Anderson transition because Vogel spirals are deterministic structures. Moreover, this transition occurs when the vector character of light is fully taken into account, in contrast to what is expected for traditional uniform random media of point-like scatterers. We show that light localization in Vogel arrays is a collective phenomenon that involves the contribution of multiple length scales. Vogel spirals are suitable photonic platforms to localize light thanks to their distinctive structural correlation properties that enable collective electromagnetic excitations with strong light-matter coupling.  Our results unveil the importance of aperiodic correlations for the engineering of photonic media with strongly enhanced light-matter coupling compared to the traditional periodic and homogeneous random media.
\end{abstract}

\pacs{Valid PACS appear here}
\keywords{Suggested keywords}
\maketitle
\section{\label{sec:level1}Introduction}\label{Introduction}
Understanding the localization of vector waves in dielectric systems provides exciting opportunities for the realization of more efficient sensors and active photonic platforms. Since the discovery by P. W. Anderson in 1958 that strong disorder can inhibit electronic transport \cite{Anderson}, the quest an optical counterpart of strong localization has motivated an intense research activity in photonic random media \cite{LagendijkToday,DiederikPhotonic}. Random lasers \cite{Cao,DiederikLaser}, multiple scattering in random media\cite{DiederikPhotonic,Pinheiro2008,Pinheiro,Bellando,SkipetrovPRL,SkipetrovPRB,Skipetrov2015,RusekPRA,RusekPRE,Lagendijk,Sheng,Maximo,Rusek2D}, local density of states modification induced by multiple scattering \cite{RiboliPRL,Tiggelen}, tuning and controlling of coupled-random modes \cite{Sgrignuoli,Vanneste,Riboli_NatMater}, and speckle pattern information decoding \cite{Fayard,Starshynov,Redding}, are some of the important results recently achieved in the field of disordered photonics. However, there is no unquestionable observation of light localization in three-dimensional (3D) uniform random systems ($i.e.$ in a full vectorial electromagnetic problem) so far \cite{Sperling,SkipetrovNJP,Scheffold,Beek2012}. The lack of materials with large enough refractive index values at optical frequencies and the presence of near-field coupling effects between scatterers in dense systems are often considered the main reasons preventing Anderson localization of light in homogeneous random media \cite{SkipetrovNJP,Naraghi}. Moreover, due to the lack of simple design rules for efficient optimization, the applications of uniform random structures to optical engineering remain quite limited. 

Aperiodic optical media, generated by simple deterministic mathematical rules, offer an alternative route to achieve light localization with respect to uniform random systems. Aperiodic deterministic systems have recently attracted significant attention in the optics and electronics communities \cite{Kohmoto,MaciaBook,Razi,DalNegroBook}. This is due, not only to their design advantages and compatibility with current nano-fabrication technologies, but also to their distinctive optical behavior \cite{DalNegro_LaserPhoton,DalNegroBook,Trevino,DalNegro_Crystals,Wang,WangEdge}. In particular, deterministic aperiodic structures display physical properties that cannot be found in either periodic or uniform random systems, such as anomalous transport \cite{DalNegroScirep,Barthelemy,Vardeny} and fractal transmission spectra \cite{Gellermann,DalNegro_PRLFibo}. 
Moreover, the tunable structural complexity of aperiodic deterministic media leads to the formation of rich spectra of resonances, called critical modes \cite{Wang,Mahler,Noh}, characterized by power-law envelope localization and multi-fractal field intensity oscillations \cite{Gellermann,DalNegro_Crystals,Wang,WangEdge}. Due to their unique functionalities, deterministic aperiodic designs have been successfully utilized in engineering applications for light emission and lasing \cite{Vardeny}, optical sensing \cite{Razi,DalNegro_LaserPhoton}, photo-detection \cite{Trevino}, nonlinear optical devices \cite{Lifshitz,Shalaev}, as well as optical imaging \cite{Huang}. 

In this manuscript we show that the large family of deterministic aperiodic Vogel arrays composed of electric dipoles can be conveniently designed to demonstrate light localization in three dimensions. We prove that a transition from diffusive to localized regimes exists in planar Vogel spiral arrays embedded in 3D space by using the dyadic Green's matrix formalism \cite{RusekPRA,RusekPRE,Lagendijk,Sheng}. The Green's matrix method has been applied to investigate Anderson localization of light in uniform disordered systems \cite{Pinheiro2008,Pinheiro,RusekPRA,RusekPRE,SkipetrovPRL,SkipetrovPRB,Skipetrov2015,Bellando,Lagendijk,Rusek2D,Maximo,Sheng} and has allowed to unveil the fundamental scattering and transport properties of aperiodic deterministic geometries \cite{DalNegro_Crystals,Wang,WangEdge}.

In this work we will focus on open planar spiral structures as they are relevant architectures for experiments and applications where light can leak out through the array plane \cite{DalNegroBook,DalNegro_LaserPhoton,Trevino,Lawrence_JAP,Pecora,LawrenceVogel}. In such situations, the electromagnetic field is not only confined in the plane of the arrays but it also leaks out to free-space according to the quality factors of the modes. We study light localization in Vogel arrays with different geometrical parameters by evaluating the Thouless conductance, the level spacing distribution, and by performing a finite-size scaling. We show the existence of different classes of localized resonances that display spatial mode profiles recently discovered across the multi-fractal band-edges of Vogel spirals using the Finite Element Method (FEM) \cite{TrevinoOpticsVogel,Liew}. Moreover, we provide a comparison with respect to open uniform random planar arrays embedded in a 3D environment for both scalar as well as vector waves unveiling the full potential of aperiodic spatial correlations for the engineering of complex photonic media with more efficient light-matter interaction. Specifically, we demonstrate that  light localization in Vogel spirals is driven by collective electromagnetic coupling effects that involve multiple length scales. For comparison, we show that vector wave localization is never achieved in planar homogeneous random systems, even when neglecting the near-field interaction term. This is shown by separately investigating the relative contributions of the different coupling terms that appear in the dyadic Green's propagator and by evaluating the Thouless conductance for sufficiently large scattering strengths. In summary,  light localization transition in Vogel spirals also occurs when the vector nature of light is taken into account, in contrast to the Anderson localization transition that it is limited to scalar waves \cite {Bellando,SkipetrovPRL}. Despite this important difference, the discovered light localization transition in Vogel spirals remarkably shares similar properties with the Anderson transition such as the crossover from level repulsion to the absence of level repulsion and the scaling properties of the Thouless conductance. Hence we conclude that  structural correlations play a crucial role in light localization in Vogel spirals as compared to uniform random systems.

This paper is organized as follows. In Sec. II we describe the Green's matrix method and the Vogel spiral photonic array. In Sec. III we present and discuss our findings whereas Sec. IV is devoted to the conclusions.
\section{\label{sec:level2} Methodology: the Vogel spiral platform and the Green's matrix formalism}\label{SecMethod}

Vogel spiral point patterns have been studied in physics, mathematics, botanics, and theoretical biology in relation to the fascinating geometrical problems offered by the field of phyllotaxis \cite{DalNegro_Crystals,Naylor,Mitchison,Adam,Liew}. This class of deterministic aperiodic media is a powerful platform for nanophotonics and nanoplasmonic applications. Polarization-insensitive light diffraction \cite{Trevino}, light-emission enhancement \cite{Lawrence_JAP,Pecora}, enhanced second-harmonic generation \cite{Capretti}, and omni-directional photonic band-gaps \cite{Pollard,Agrawal} are some of them. Vogel spiral geometries are characterized by diffuse scattering spectra like uniform random media but with circularly symmetric scattering rings that can be easily controlled by the spiral geometry \cite{DalNegro_Crystals,TrevinoOpticsVogel} (see also Supplementary Material). By using simple generation rules, particle arrays with Vogel spiral geometry can be easily designed to produce a very rich structural complexity best described by multi-fractal geometry \cite{TrevinoOpticsVogel}. Moreover, Vogel spirals support distinctive scattering resonances that have been shown to encode well-defined numerical sequences in the orbital angular momentum of light, which have a great potential for device applications to singular optics and optical cryptography \cite{DalNegro_OptExpress,Simon}.
 
Vogel spiral arrays are defined in polar coordinates ($r,\theta$) by the following parametric equations:
\begin{eqnarray}\label{1}
&r_n=a_0\sqrt{n}\\
&\theta_n=n\alpha\label{2}
\end{eqnarray}
where $n = 0, 1, 2, . . .$ is an integer, $a_0$ is a positive constant called scaling factor, and $\alpha$ is an irrational number, known as the divergence angle \cite{Adam}. This angle specifies the constant aperture between successive point particles in the array \cite{Naylor}. Since the divergence angle is an irrational number, Vogel spiral point patterns lack both translational and rotational symmetry. The divergence angle ($\alpha^{\circ}$, in degrees) can be specified by  the choice of an irrational number $\xi$ according to the relationship $\alpha^{\circ}$=360$^{\circ}-frac(\xi)\times{360^{\circ}}$ where $frac(\xi)$ denotes the fractional part of $\xi$. Vogel spirals with remarkably different structural properties can be obtained simply by selecting different values for the irrational number $\xi$. For instance, when $\xi$ is equal to the golden mean $\xi$=$(1+\sqrt{5})/2$ the corresponding divergence angle $\alpha\sim{137.508^{\circ}}$ is called the ``golden angle" while the resulting Vogel spiral structure is called the golden angle spiral, or GA spiral. This provides opportunities to tailor different degrees of aperiodic structural order in a very efficient way \cite{DalNegro_Crystals} (see Supplementary Material for more details).

In this work, we primarily focus on four different types of Vogel spiral arrays introduced in Refs.\cite{DalNegro_Crystals,DalNegro_OptExpress,LawrenceVogel}, which are called GA-spiral, $\tau$-spiral, $\pi$-spiral, and $\mu$-spiral. They are generated according to Eq.(\ref{1}) and Eq.(\ref{2}) choosing the values  $\xi$=$(1+\sqrt{5})/2$, $\xi$=$(5+\sqrt{29})/2$, $\xi$=$\pi$, and $\xi$=$(2+\sqrt{8})/2$, respectively. The $\pi$-spiral exhibits the lowest degree of structural order, followed by the $\mu$-spiral, the $\tau$-spiral, and the GA-spiral \cite{DalNegro_Crystals} (see also Supplementary Material). This ordering reflects the smallest number of convergents ($i.e.$ rational approximations) necessary to approximate the irrational number $\xi$ in continued fractions at any level of accuracy \cite{DalNegro_Crystals,DalNegro_OptExpress}. 

We now investigate the spectral and wave localization properties of Vogel spirals using the Green's matrix method. This approach provides access to all the scattering resonances of a system composed of vector electric dipoles in vacuum and accounts for all the multiple scattering orders, so that multiple scattering process is treated exactly. In addition, this method allows for a full description of scattering resonances of a large-scale 
structures at a relatively low computational cost if compared to traditional numerical methods such as Finite Difference Time Domain (FDTD) or FEM techniques. Each scatterer is characterized by a Breit-Wigner resonance at frequency $\omega_0$ and width $\Gamma_0$ ($\Gamma_0$$\ll\omega_0$). The quasi-modes of this scattering system can be identified with the eigenvectors of the Green's matrix $\overleftrightarrow{G}$ which, for $N$ vector dipoles, is a $3N\times{3N}$ matrix with components \cite{SkipetrovPRL}:
\begin{equation}\label{Green}
G_{ij}=i\left(\delta_{ij}+\tilde{G}_{ij}\right)
\end{equation}
$\tilde{G}_{ij}$ has the form:
\begin{eqnarray}\label{GreenOur}
\begin{aligned}
\tilde{G}_{ij}=&\frac{3}{2}\left(1-\delta_{ij}\right)\frac{e^{ik_0r_{ij}}}{ik_0r_{ij}}\Biggl\{\Bigl[\bold{U}-\hat{\bold{r}}_{ij}\hat{\bold{r}}_{ij}\Bigr]\\
&- \Bigl(\bold{U}-3\hat{\bold{r}}_{ij}\hat{\bold{r}}_{ij}\Bigr)\left[\frac{1}{(k_0r_{ij})^2}+\frac{1}{ik_0r_{ij}}\right]\Biggr\}
\end{aligned}
\end{eqnarray}
when $i\neq j$ and $0$ for $i=j$. $k_0$ is the wavevector of light, the integer indexes $i, j \in 1,\cdots,N$ refer to different particles, $\textbf{U}$ is the 3$\times$3 identity matrix, $\hat{\bold{r}}_{ij}$ is the unit vector position from the $i$-th and $j$-th scatter while $r_{ij}$ identifies its magnitude. This method is suitable not only for the study of atomic clouds, as atoms are perfect dipoles, but also provides fundamental insights into the physics of periodic, aperiodic, and uniform random systems of small scattering particles \cite{Pinheiro2008,Pinheiro,Bellando,SkipetrovPRL,SkipetrovPRB,Skipetrov2015,RusekPRA,RusekPRE,Lagendijk,Sheng,Maximo,Rusek2D,DalNegro_Crystals,Wang,WangEdge}. The Green's matrix (\ref{Green}) is a non-Hermitian matrix. As a consequence, it has complex eigenvalues $\Lambda_n$ ($n \in 1,2,\cdots, 3N$) \cite{RusekPRA,RusekPRE}. The real and the imaginary part of $\Lambda_n$ are related to the detuned scattering frequency $(\omega_0-\omega_n)$ and to the scattering resonance decay $\Gamma_n$ both normalized with respect to the resonant width $\Gamma_0$ of an isolated dipole \cite{SkipetrovPRL,SkipetrovPRB,Skipetrov2015}. In the following, we define $\hat{\omega}_n$=$(\omega_0-\omega_n)/\Gamma_0$. 

In order to establish light localization as a breakdown of photon diffusion, we have analyzed two parameters. The first parameter characterizes the degree of spectral overlap between different optical resonances and it is called Thouless conductance $g$ \cite{Thouless,Edwards}. The second parameter quantifies the sensitivity/insensitivity of scattering resonances with respect to  a perturbation of the system boundary conditions and it is known as the $\beta$ parameter. In order to demonstrate light localization we have applied two criteria. First, the $g$ conductance, which is proportional to the scattering mean free path of the system, must decrease when increasing the scattering strength, $i.e.$ increasing the optical density $\rho\lambda^2$. Here $\rho$ is the number of particles per unit area while $\lambda$ is the optical wavelength. Second, the scaling of the $\beta$ parameter with respect to the logarithmic conductance ($\beta$=$\beta[\ln(g)]$) must show a critical point $q_c$=$\ln(g_c)$ at which $\beta$ vanishes, $i.e.$ the Thouless conductance does not depend on the system size $L$ \cite{Abrahams,Abrahams50years}.
 
Within the Green's matrix formalism, the Thouless conductance is defined as the ratio of the dimensionless lifetime $(\delta\omega)^{-1}$=$1/\Im[\Lambda_n]$ to the spacing of nearest dimensionless resonant frequencies $\Delta\omega$=$\Re[\Lambda_n]-\Re[\Lambda_{n-1}]$ \cite{SkipetrovPRL}. In order to study the behavior of $g$ as a function of the resonance frequencies, we have  subdivided, for each value of the scattering strength $\rho\lambda^2$,  the range of resonance frequencies in $300$ equispaced intervals. This allows to consider the average value of $g$ within each subinterval and to obtain its frequency dependence by plotting the average values associated to each subinterval. The Thouless conductance $g$ can be written in terms of the eigenvalues of the Green's matrix as
\begin{equation}\label{Thouless}
g=g(\omega)=\frac{\overline{\delta\omega}}{\overline{\Delta\omega}}=\frac{(\overline{1/\Im[\Lambda_n]})^{-1}}{\overline{\Re[\Lambda_n]-\Re[\Lambda_{n-1}]}}
\end{equation}
where $\overline{\{\cdots\}}$ indicates the average of $g$ over each frequency subinterval. The frequency $\omega$ is the central frequency of each subinterval used to sample the $\Re[\Lambda_n]$ axes. Differently from the uniform random scenario \cite{Pinheiro2008,Pinheiro,Bellando,SkipetrovPRL,SkipetrovPRB,Skipetrov2015,RusekPRA,RusekPRE,Lagendijk,Sheng,Maximo,Rusek2D,Goetschy}, we do not perform any average with respect to different geometry configurations because Vogel spirals are deterministic structures.
\section{\label{sec:level4} Results and Discussions}\label{Results}
\begin{figure*}[h!]
\includegraphics[width=12cm]{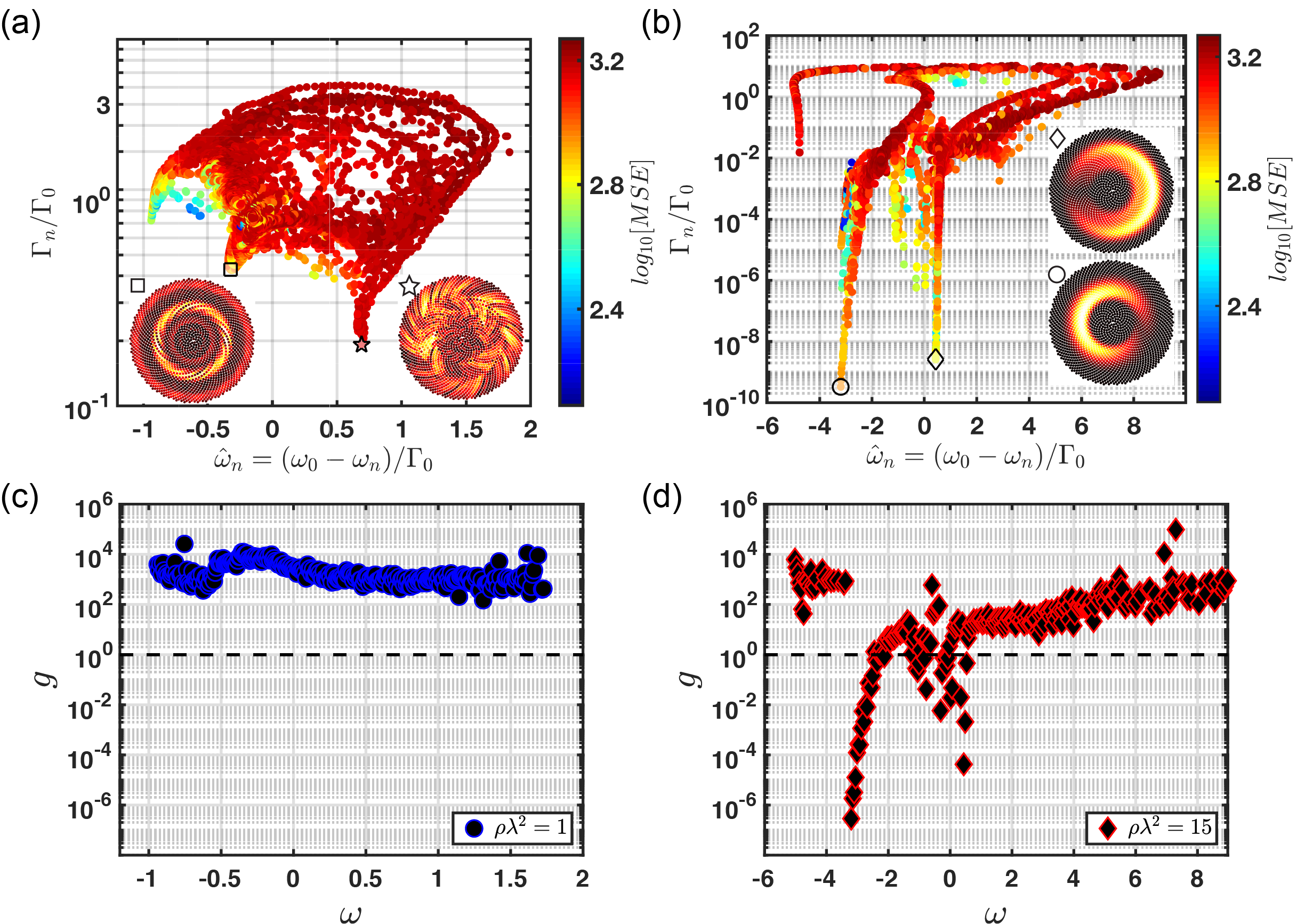}
\caption{Eigenvalues of the Green's matrix (\ref{Green}) are shown by points on the complex plane for 2000 electric point dipoles arranged in a GA Vogel spiral geometry. Panel (a) and (b) refer to an optical density of 1 and 15, respectively. Scattering resonances with very small decay rates ($\Im[\Lambda_n]=\Gamma_n/\Gamma_0\ll1$) appear only when $\rho\lambda^2$=15. The data are color-coded according to the $\log_{10}$ values of the MSE. Insets: spatial profiles of representative quasi-modes. Panel (c) and (d) show the frequency dependence of the Thouless conductance $g$ when $\rho\lambda^2$ is equal to 1 and 15, respectively. The dashed-black lines identify the threshold of the diffusion-localization transition.}
\label{Fig1}
\end{figure*}

We will first consider the case of $N=2000$ electric vector dipoles arranged in a GA Vogel spiral configuration. The 3N$\times$3N Green's matrix (\ref{Green}) is diagonalized numerically and the Thouless conductance $g$, defined by Eq.(\ref{Thouless}), is calculated as a function of the frequency $\omega$ for different values of $\rho\lambda^2$. Fig.\ref{Fig1} panel (a) and (b) show the distribution of the resonant complex poles $\Lambda_n$, color-coded according to the $\log_{10}$ values of the Mode Spatial Extent (MSE), when the optical density is set equal to $1$ and to $15$, respectively. The MSE parameter characterizes the spatial extent of a photonic mode \cite{Sgrignuoli}. 

At low optical density ($\rho\lambda^2$=1), the system is in the delocalized regime. The complex eigenvalue distribution does not show the formation of any long-lived resonances with $\Gamma_n/\Gamma_0\ll1$. Consistently, the spatial profiles of the modes in this regime are delocalized across the array. For example, two representative eigenvectors that correspond to the smallest decay rates around $\hat{\omega}_n$$\sim-$0.3 (white-square marker) and $\hat{\omega}_n$$\sim$0.8 (white-pentagram marker), are shown in the inset of Fig.\ref{Fig1} (a). Expectedly, we found that the Thouless conductance is always larger than one in this case, as shown in Fig.\ref{Fig1} (c).

However, at large optical density ($\rho\lambda^2=$15), the situation is completely different. Long-lived resonances appear and a significant fraction of the complex eigenvalues of the Green's matrix have a very small decay rate ($\Gamma_n/\Gamma_0\ll1$). For comparison, no such long-lived resonances appear in uniform random media when the vector nature of light is taken into account (see Appendix \ref{AppxA} and Refs.\cite{SkipetrovPRL,Bellando,Maximo} for more details). We show the spatial profiles of two representative eigenvectors in the inset of Fig.\ref{Fig1} (b) and we report in Fig.\ref{Fig1} (d) the Thouless conductance as a function of frequency. These findings clearly demonstrate that the system reached the localization regime at large optical density, namely the eigenvectors are radially confined and $g(\omega)<1$. We notice that the long-lived resonances shown in Fig.\ref{Fig1} (b) are clustered around two ``tail regions" that appear at the frequency positions where $g$ becomes lower than one (see Fig.\ref{Fig1} (d)). 

Two other important features arise at sufficiently large optical density: the existence of a spectral gap region and the absence of sub-radiant ``dark" states, also called proximity resonances, in the complex distribution of the eigenvalues \cite{RusekPRA,Heller,DalNegro_Crystals}. Proximity resonances are sub-radiant states localized around pairs of scatterers and can be identified in random systems by their typical spiral distributions in the complex eigenvalue plane  and are characterized by MSE=2 \cite{RusekPRA,Bellando,Goetschy}. The absence of proximity resonances in Vogel spiral systems was originally reported in Ref.\cite{DalNegro_Crystals} and attributed to the more regular structure of Vogel spirals compared to random media. This can be understood based on the fact that, for a given optical density, the first-neighbor distance of the particles is, on average, larger in the case of Vogel spirals. Indeed, we have previously shown that the probability distribution of first-neighbor distances is non-Gaussian for Vogel spirals and characterized by long tails \cite{DalNegro_Crystals,TrevinoOpticsVogel,DalNegroBook,Bettotti}.  More specifically, the mean value of the first-neighbor distances of the GA-spiral is $\overline{\delta}^{1st}$=$1.70\pm0.02$ (in units of the scaling factor $a_{0}$). In contrast, uniform random point patterns, characterized by the same density, are characterized by a Poissonian first-neighbor distribution \cite{Illian} with larger fluctuations: $<\overline{\delta}^{1st}>_e$=$0.89\pm0.47$. 
Here $<\cdots>_e$ indicates the average over two hundreds different point pattern realizations. Therefore, for a given optical density the probability of observing two very close particles is much larger for the uniform random patterns (see Appendix \ref{AppxB} for more details). Interestingly, these fluctuations increase up to almost $20\%$ in the $\pi$-spiral configuration, which in fact is the most disordered Vogel spiral considered in this work. The lack of significant contributions from the sub-radiant resonances in Vogel spiral has profound consequences for light localization and simplify considerably the analysis of $g$ and the $\beta$-scaling compared to  uniform random systems where the proximity resonances need to be carefully removed \cite{SkipetrovPRL,SkipetrovPRB,Bellando}.

\begin{figure*}[h!]
\includegraphics[width=12cm]{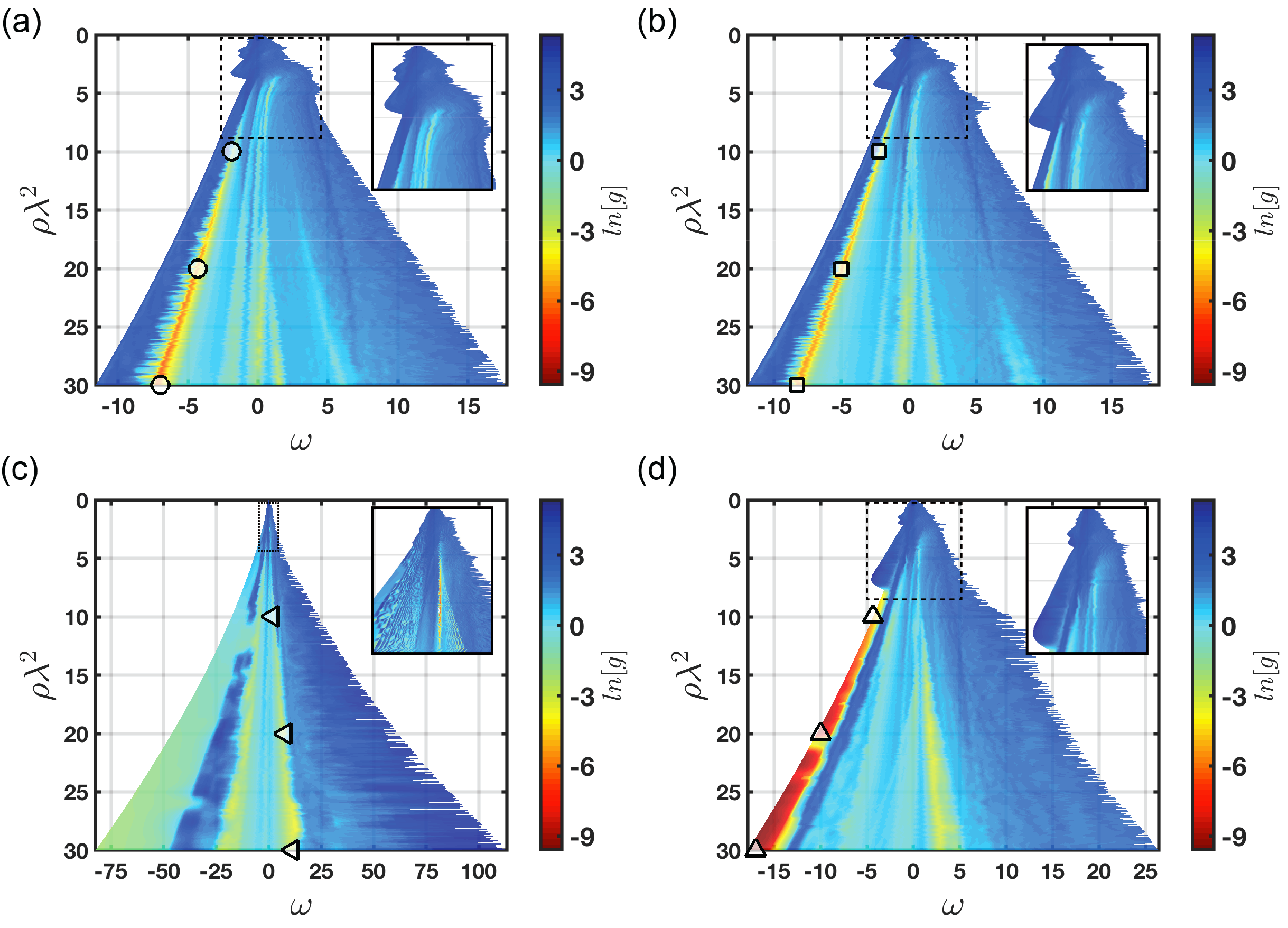}
\caption{Highly resolved maps of the logarithmic values of the averaged Thouless conductance are evaluated for different values of the optical density $\rho\lambda^2$ as a function of $\omega$. The data of panels (a-d) are color-coded according to $\ln[g]$ and refer to the GA-spiral, $\tau$-spiral, $\pi$-spiral, and $\mu$-spiral, respectively. Each Thouless conductance $g(\omega)$ is characterized by 300 points. These maps are calculated in the range $\rho\lambda^2$=[0.1,30] with a resolution of $0.1$. Insets: enlarged view of the threshold region for the diffusion-localization transition. Green-red-yellow features identify the appearance of localized resonances that follow clear dispersion branches with respect to $\omega$. Different markers identify the classes of localized resonances that produce the stronger localization feature in the considered Vogel spirals.}
\label{Fig2}
\end{figure*}

In order to gain more insights on the localization transition we study the logarithm of the averaged Thouless conductance for different values of the optical density (starting from 0.1 up to 30 with a resolution of  $\rho\lambda^2$=0.1) as a function of $\omega$. In this way, highly resolved maps of the quantity $\ln[g]=\ln[g(\omega,\rho\lambda^2)]$ can be obtained. The results of this analysis are summarized in Fig.\ref{Fig2} (a-d) for the GA, $\tau$, $\pi$, and $\mu$ spirals, respectively. The data are color-coded according to the logarithmic values of the Thouless conductance. The diffusion-localization threshold is defined according to $\ln[g(\omega,\rho\lambda^2]=0$ and it is identified by the cyan color. Insets display enlarged views of the threshold region for the diffusion-localization transition. Localization begins to take place at $\rho\lambda^2\sim3.5$ for all the geometries except for the more disordered $\pi$-spiral configuration, whose threshold occurs at $\rho\lambda^2\sim2$. While this analysis focused on spirals with $N$=$2000$, we have numerically verified that the the results are robust with respect to system size ($N$=$500-4000$) and the frequency resolution used in the computation of the Thouless conductance $g$. 

The appearance of localized resonances, identified by the green-red-yellow features in Fig.\ref{Fig2}, shows a clear dispersion branch with respect to the frequency $\omega$ in all the investigated geometries. These features cannot be obtained in a uniform random medium where the attainable value of the Thouless conductance are always larger than one \cite{Bellando,SkipetrovPRL}. (See also Appendix \ref{AppxA} for a detailed comparison with respect to the vector and the scalar model of uniform random planar arrays embedded in a 3D space with the same optical density of Vogel spirals). In Vogel spirals different classes of localized resonances are clearly visible in Fig.\ref{Fig2}. 

\begin{figure*}[h!]
\includegraphics[width=12cm]{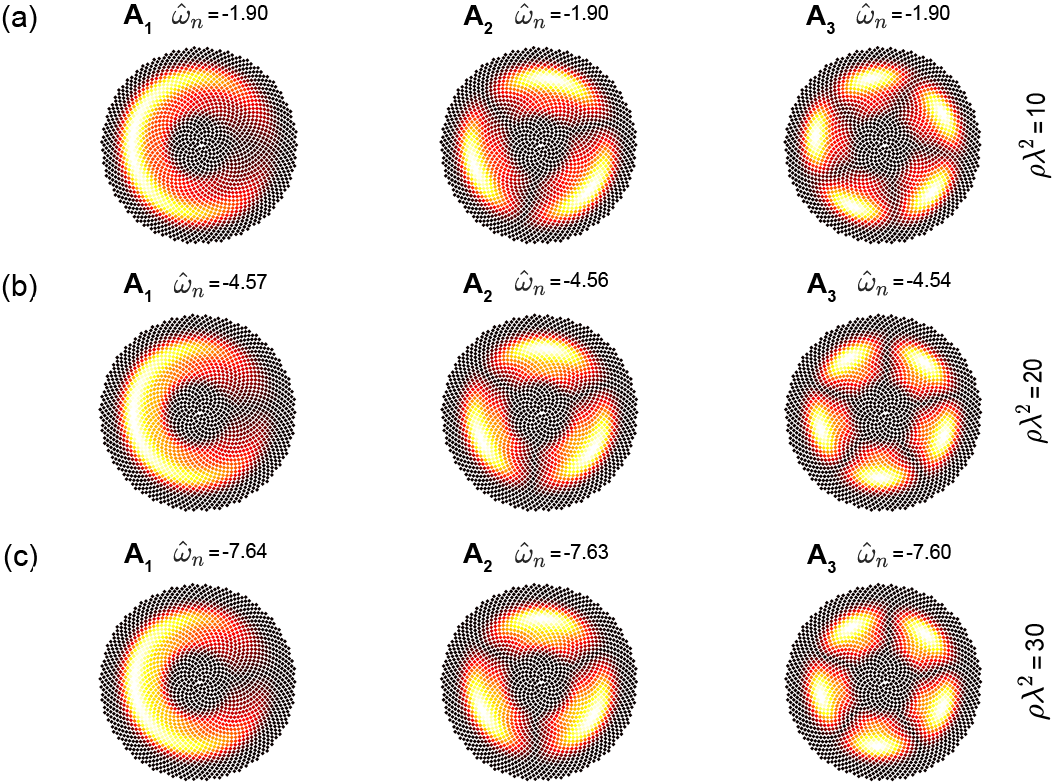}
\caption{Representative spatial distributions of the Green's matrix eigenvectors that belong to the class of scattering resonances that produce the stronger localization feature in the the GA-spiral. A$_1$, A$_2$ , and A$_3$ identify the first three resonances with the lower decay rates( $\Gamma_n/\Gamma_0\ll1$). Panels (a-c) show these quasi-modes when the optical density $\rho\lambda^2$ is fixed to 10, 20, 30, respectively. The spectral positions of these quasi-modes are identified by the white-circle markers in Fig.\ref{2} (a).}
\label{Fig3}
\end{figure*}

To achieve a deeper understanding on the nature of the resonances that belong to different dispersion branches, we have systematically analyzed the spatial distributions of few representative examples identified by the markers shown in Fig.\ref{Fig2}. These markers identify the behavior of the class of scattering resonances that produce the stronger localization feature in the considered Vogel spirals. We first focus on the type of resonances highlighted with the white-circle markers in Fig.\ref{Fig2} (a). The spatial distributions of eigenvectors of the Green's matrix corresponding to the three resonances with the lower decay rates are labeled in Fig.\ref{Fig3} as $A_1$, $A_2$, and A$_3$, respectively. In Fig.\ref{Fig3} (a-c) the optical density $\rho\lambda^2$ is fixed to 10, 20, and 30, respectively. For each one of them, the frequency of the scattering resonance $\hat{\omega}_n$ is also indicated. It is clearly shown that exactly the same spatial profile is retrieved when scanning along the dispersion branches for all the three resonances $A_1$, $A_2$, and $A_3$. The effect of increasing the optical density $\rho\lambda^2$ is simply to produce a frequency shift in the complex scattering plane. 
Interestingly, we notice that the spatial profiles of the scattering resonances shown in Fig.\ref{Fig3} agree very well with what has been previously reported based on the FEM method \cite{Liew,TrevinoOpticsVogel,Bettotti}, demonstrating the power of the more efficient Green's matrix approach. In our previous numerical studies we discovered that the localized modes of Vogel spirals have a quality factor that scales linearly with the frequency, which allowed us to classify them into different classes \cite{Liew,TrevinoOpticsVogel,Bettotti}. The modes belonging to the same class have similar spatial patterns and each one of them has a degenerate counterpart characterized by a complementary spatial profile. We now report a complete classification of the Vogel spiral modes based on the more systematic Green's matrix analysis that provides access to all the localized modes that exist for a given optical density in the structure. As an example, the three types of resonances shown in  Fig.\ref{Fig3} have exactly the same spatial profiles that correspond to band-edge modes of class A, as defined in Refs.\cite{Liew,TrevinoOpticsVogel,Bettotti}. Moreover, also the degenerate modes of $A_1$, $A_2$, and $A_3$ can be identified by using the Green's matrix formalism. They occur exactly at the same $\hat{\omega}_n$ and they are characterized by a complementary spatial profile. This comparison demonstrates also that light localization in Vogel spirals is produced at the band-edge due to the strongly-fluctuating (multi-fractal) dispersion in the density of states \cite{TrevinoOpticsVogel}. Exactly the same conclusions are obtained for the $\tau$-spiral, $\pi$-spiral, and $\mu$-spiral (see Appendix \ref{AppxC} for more details). In summary, the aperiodicity and the distinctively structural correlation properties of Vogel spirals produce different classes of localized band-edge modes that can be successfully identified and characterized using the Green's matrix spectral method, demonstrating an additional methodological progress that can be achieved using this method. 

\begin{figure*}[h!]
\includegraphics[width=12cm]{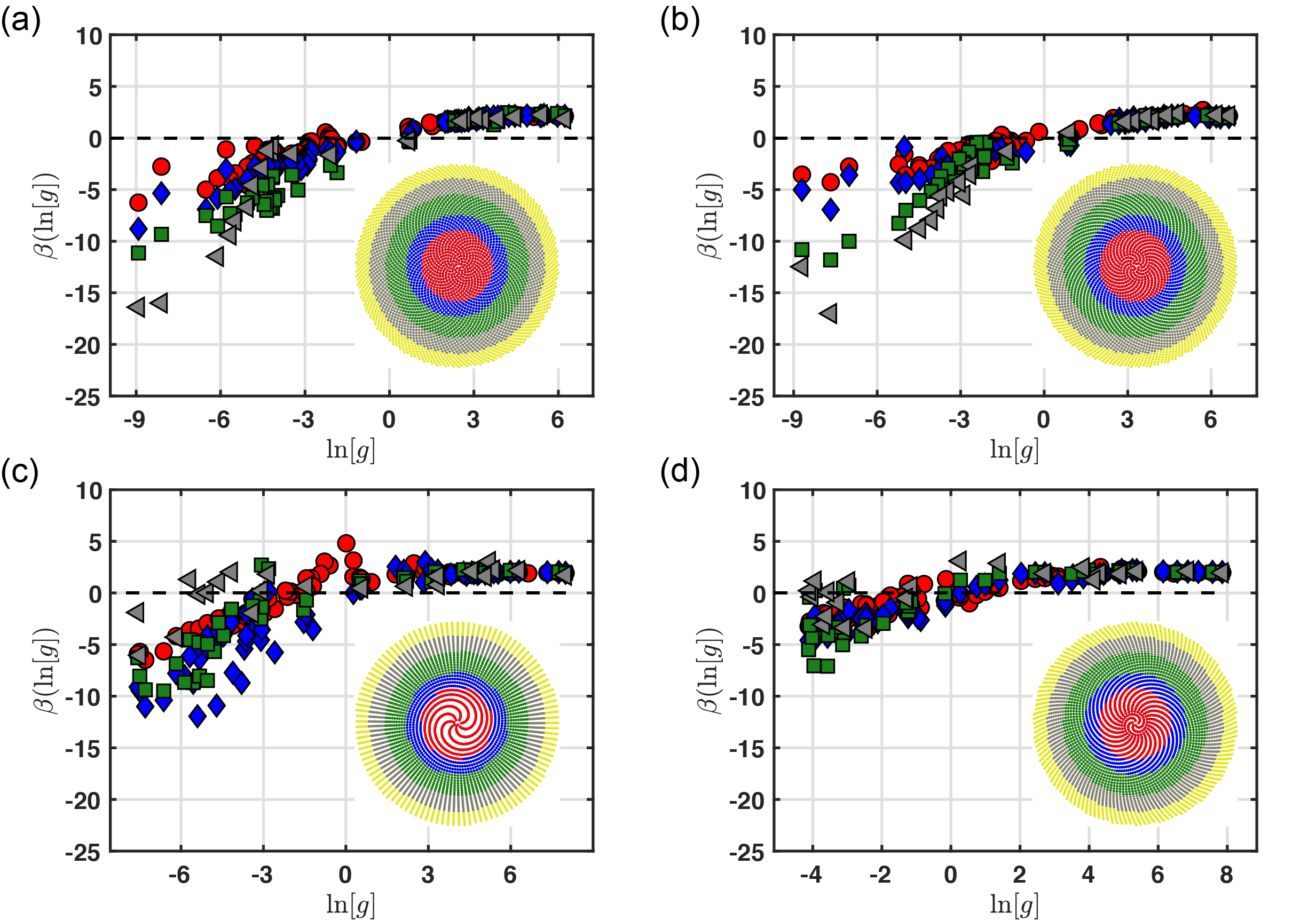}
\caption{The $\beta$-scaling analysis is performed by increasing the number of scatters from $N$=$500$ up to $N$=$4000$. Panels (a-d) display the results of this scaling for the GA-spiral, $\tau$-spiral, $\pi$-spiral, and $\mu$-spiral, respectively. Different markers correspond to calculations of the $\beta$ function for all the possible combinations of $N$. Red-circles, blue-diamonds, olive-squares, and gray-triangles determine the combination of $N$=$500$ versus $N$=$1; 2; 3; 4\times10^3$, $N$=$1000$ versus $N$=$2; 3; 4\times10^3$, $N$=$2000$ versus $N$=$3; 4\times10^3$, and $N$=$3000$ with respect to $N$=$4\times10^3$, respectively.The dashed-black lines identify the $\beta$-parameter transition threshold. Insets: set of different Vogel spirals generated by increasing the number of scatters: $N$=4000 (yellow), $N$=3000 (grey), $N$=2000 (olive), $N$=1000 (blue), and $N$=500 (red).}
\label{Fig4}
\end{figure*}

We will now address the scaling analysis of the localization transition in Vogel spirals. This analysis is based on the assumption that a phase transition between localized and diffusive states exists only in a three dimensional scenario, while the system is always insulating in lower dimensions \cite{Abrahams}. Therefore, the diffusive and localized states are separated by a critical point, called the ''mobility edge". The scaling analysis is characterized by only one parameter, the Thouless conductance $g$. According to the scaling theory of localization, the dependence of the conductance on the system size can be described by the $\beta$-function \cite{Abrahams}:
\begin{equation}\label{beta} 
\beta(\ln[g])=\frac{d\ln[g]}{d\ln[L]}
\end{equation}
where $L$ is the product of the wavevector $k_0$ and the system size $R$, which is the maximum radial coordinate of the spiral (see insets of Fig.\ref{Fig4}). Eq.(\ref{beta}) assumes that Thouless conductance $g$ is a continuous and monotonic function of $L$. Fig.\ref{Fig4} (a-d) display the results of the scaling analysis applied to GA-spiral, $\tau$-spiral, $\pi$-spiral  and $\mu$-spiral, respectively, by increasing the number of scatters from $N$=500 up to $N$=4000. 
In each analyzed Vogel spiral geometry, a clear transition from $\beta>0$, where the conductance grows with $L$ and the eigenstates are delocalized, to $\beta<$0, where the conductance decreases with $L$ manifesting the presence of localized eigenstates, occurs. We discovered that the scaling of the $\beta$ parameter with $L$ is characterized by a critical points $q_c$=$\ln[g_c]$  at which $\beta(q_c)$=0, indicating that $g_c$ becomes independent of the system size. At this point, the trends of the $\beta$ parameter do not show any significant deviations with the system size, as shown in Fig.\ref{Fig4} for Vogel spirals with $N$=500$-$4000. This demonstrates that our data are compatible with the single-parameter scaling hypothesis expressed by Eq.(\ref{beta}). 

 Our results demonstrate, for the first time, a light localization transition supported by open Vogel spirals planar arrays embedded in three dimensions. This phenomenon cannot occur in traditional uniform random media when the vector nature of light is taken into account within the Green's matrix formalism (see Refs. \cite{SkipetrovPRL,Bellando,Maximo} and Appendices \ref{AppxA} and \ref{AppxD} for more details). These results put into evidence the main difference between the light localization transition in Vogel spirals and the Anderson localization transition, which occurs only in the scalar approximation for point-like electric dipoles \cite{Bellando,SkipetrovPRL}. However, the light localization transition in Vogel spirals remarkably shares similar properties with the Anderson transition such as the crossover from level repulsion to the absence of level repulsion (as demonstrated in Appendix \ref{AppxE}) and the scaling properties of the Thouless conductance. It is important to note that in our study, although the dipoles are arranged in planar Vogel spiral arrays, the electromagnetic field is not only confined in the plane but it also leaks out to free-space according to the quality factors of the modes. Therefore such systems are truly open scattering 3D systems. On the contrary, in a strictly two-dimensional (2D) problem the electromagnetic field is described by a 2D Green's function ($i.e.$ Hankel function) and it is uniform along the vertical z-coordinate (see $e.g.$  \cite{Rusek2D,RiboliPRL,Maximo,Sheng}).

\begin{figure*}[h!]
\includegraphics[width=12cm]{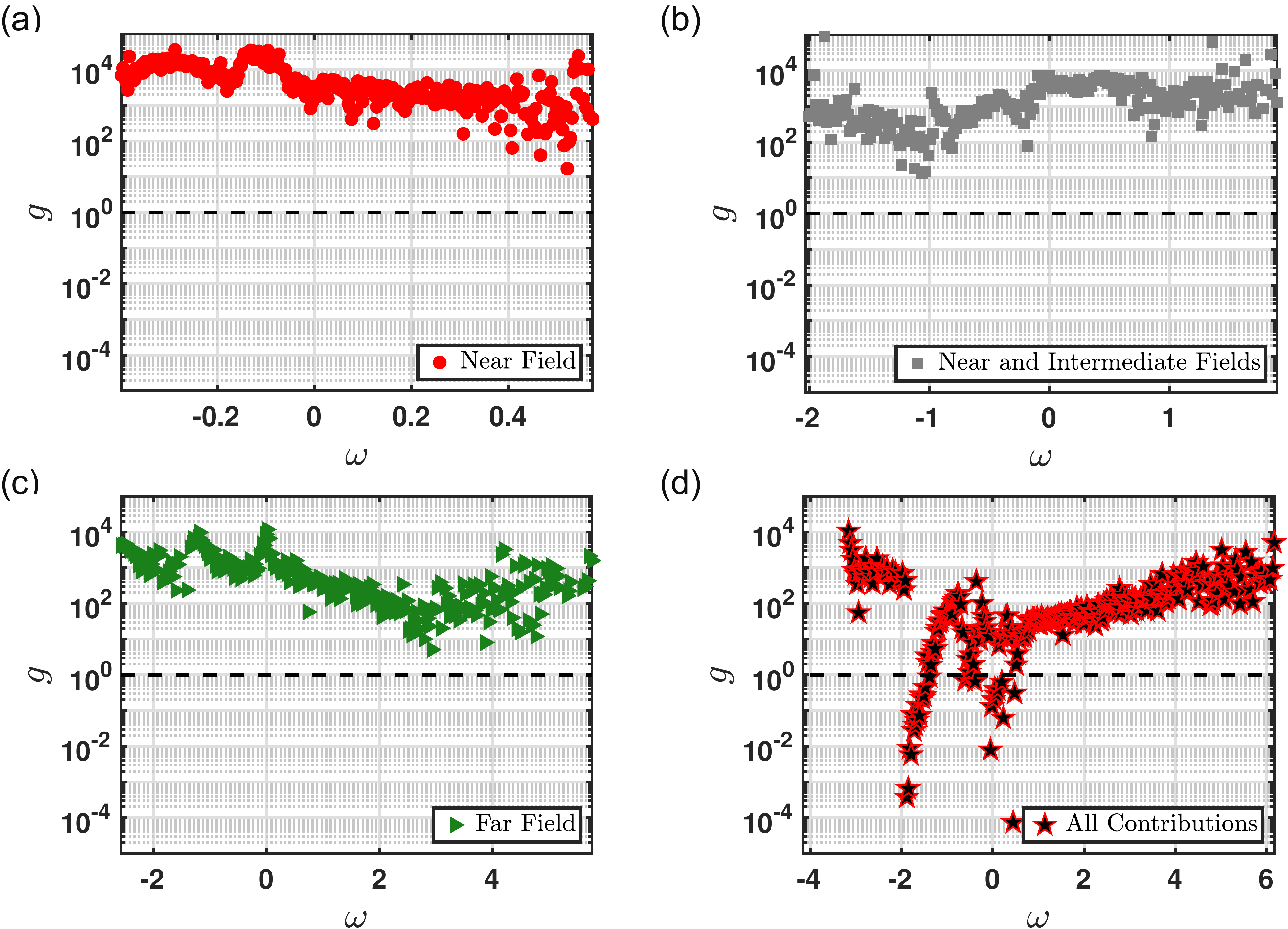}
\caption{Panels (a-d) display, in a semi-log-y scale, the frequency dependence of $g$ after diagonalizing the 3$N$$\times$3$N$ Green's matrix associated to only the near-field term, the near-field term plus the intermediate-field contribution, the far-field term only, and all the coupling contributions, respectively. These data refer to the GA-spiral when the optical density $\rho\lambda^2$=10. The dashed-black lines identify the threshold of the diffusion-localization transition $g$=1. Similar results are obtained for all the other investigate Vogel spirals and compared with uniform random media in Appendix \ref{AppxD}.}
\label{Fig5}
\end{figure*}

In order to investigate the role of cooperative effects in the light localization of Vogel spirals we have decomposed the Green's matrix of Eq.(\ref{GreenOur}) into the sum of three coupling terms. Each term describes different electromagnetic coupling regimes proportional to $1/r_{ij}$, $1/r^2_{ij}$, and $1/r^3_{ij}$, corresponding to long-range, intermediate-range, and short-range electromagnetic interactions, respectively. We separately investigated these different contributions of the dyadic Green's propagator and, for each one of them and for their different combinations, we evaluated the Thouless conductance for an optical density $\rho\lambda^2=$10. Fig.\ref{Fig5} summarizes our results for the case of a GA-spiral (similar results are obtained for all the other investigated Vogel spirals and compared with uniform random media in Appendix \ref{AppxD}). Panels (a-d) show the frequency dependence of the Thouless conductance $g$ obtained by using Eq.(\ref{Thouless}) after diagonalizing the 3N$\times$3N Green's matrix associated to only the near-field term, the near-field term plus the intermediate-field contribution, the far-field term only, and all the coupling contributions, respectively. Light localization, characterized by $g<$1, occurs only when all the coupling terms, including the near-field regime, are simultaneously taken into account. Therefore, our results demonstrate for the first time that light localization in Vogel spirals results from a collective coupling effect that involves multiple length scales. Remarkably, we also demonstrate that vector wave localization is never achieved in uniform random systems with a planar support, even neglecting the near-field interaction term (see Appendices \ref{AppxA} and \ref{AppxD} for more details).

\begin{figure*}[h!]
\includegraphics[width=12cm]{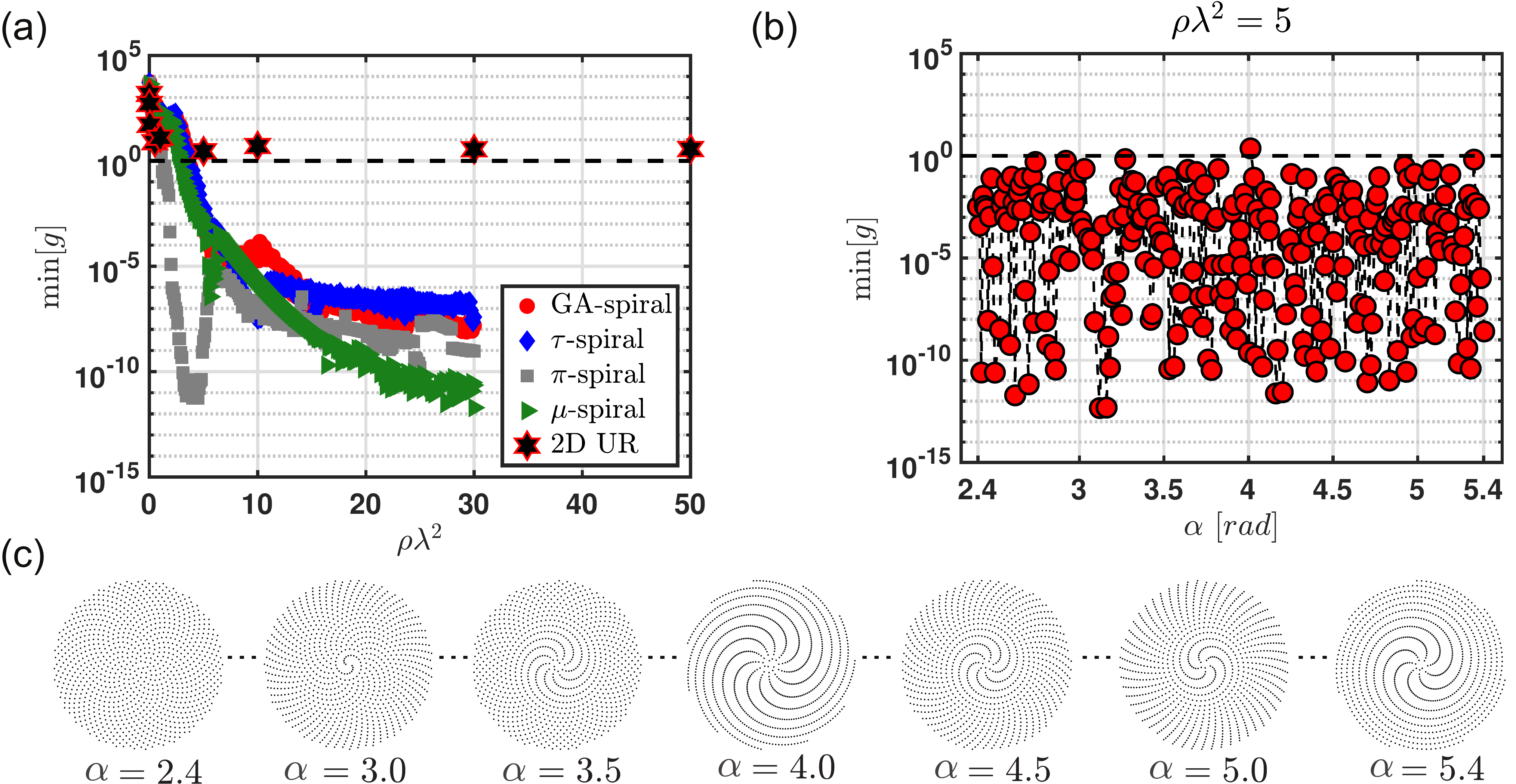}
\caption{(a) The minimum value of the Thouless conductance as a function of $\rho\lambda^2$ is  reported in a semi-log-y scale. Different markers identify the different analyzed structures. Red-circle markers, blue-diamond markers, gray-square markers, olive-triangle markers, and red/black-hexagram markers refer to the GA-spiral, $\tau$-spiral, $\pi$-spiral, $\mu$-spiral, and to the uniform random configuration ($i.e.$ 2D UR), respectively. 10 different disorder realizations are considered for the 2D UR analysis. (b) $\min[g]$ behavior as a function of the divergence angle $\alpha$ (expressed in radiant) when $\rho\lambda^2$=5. 300 Vogel spirals, with remarkably different structural correlations, are generated between $\alpha=2.4$ rad and $\alpha\sim5.4$ rad. Some representative Vogel spiral geometries are shown in panel (c). The dashed-black lines identify the threshold of the diffusion-localization transition $g$=1.}
\label{Fig6}
\end{figure*}

The effect of the optical density on the minimum value of the Thouless conductance $g$ is illustrated in Fig.\ref{Fig6}(a) where we also compare with the case of planar uniform random media, referred as 2D UR in the legend. All the structures have $N$=2000 interacting particles and the random system's results are averaged over 10 different realizations. Moreover, in order to eliminate the contribution of proximity resonances from the analysis of the random configuration, we have carefully neglected the resonances with $MSE$=2 \cite{RusekPRA,Bellando,Goetschy}. Fig.\ref{Fig6}(a) shows that light localization never appears in uniform random arrays. 
This analysis is performed for different values of $\rho\lambda^2$ up to 50. In contrast, all the Vogel spirals exhibit light localization starting from a threshold value of $\rho\lambda^2\geq{2}$, as previously discussed. The $\pi$-spiral configuration, whose geometry is the most disordered, displays the lowest localization threshold as well as the minimum $g$ value. In order to generalize our findings to a much larger set of Vogel spirals we compute in Fig.\ref{Fig6} (b) the minimum value of $g$ at optical density $\rho\lambda^2$=5 for 300 different Vogel spirals obtained by continuously varying the polar divergence angle $\alpha$ defined in Eq.(\ref{2}). All these structures are generated with a divergence angle that linearly interpolates between the GA-spiral and the $\pi$-spiral. Some representative geometries are shown in Fig.\ref{Fig6} (c). These data demonstrate that vector wave localization is a very robust feature of Vogel spiral arrays that can be achieved for many different choices of the divergence angle $\alpha$. The results of our paper clearly establish the relevance of controllable aperiodic correlations for the engineering of photonic scattering platforms with strong light-matter interaction.
\section{\label{sec:level5} Conclusions}

In summary, we have demonstrated a light localization transition supported by Vogel spiral planar arrays embedded in three dimensions by means of the dyadic Green's matrix method. Specifically, a clear transition from the diffusive to the localized regime, different from the Anderson localization transition in three dimensions, is discovered by evaluating the Thouless conductance, the level spacing statistics, and by performing a systematic finite-size scaling analysis. Different classes of localized modes of Vogel spirals with distinctive spatial distributions have been identified and analyzed. By decomposing the dyadic field propagator in its different components we show that light localization in Vogel arrays originates from collective electromagnetic coupling involving the contributions of multiple length scales. All these effects do not occur in traditional uniform random media. Our results unveil the importance of structural correlations in deterministic aperiodic photonic media for the design of localized states with strongly enhanced light-matter interactions. In addition, our findings may open new vistas for the engineering of mesoscopic transport and localization phenomena and should encourage deeper investigations of photonic devices based on deterministic aperiodic architectures. 
\begin{acknowledgments}
This research was sponsored by the Army Research Laboratory and was accomplished under Cooperative Agreement Number W911NF-12-2-0023. The views and conclusions contained in this document are those of the authors and should not be interpreted as representing the official policies, either expressed or implied, of the Army Research Laboratory or the U.S. Government. The U.S. Government is authorized to reproduce and distribute reprints for Government purposes notwithstanding any copyright notation herein. F.A.P thanks the Brazilian agencies CNPq, CAPES, and FAPERJ as well as The Royal Society-Newton Advanced Fellowship (Grant No. NA150208) for financial support.
\end{acknowledgments}
\appendix
\section{Planar uniform random configuration}\label{AppxA}

The relevant features of light localization properties of uniform random arrays are presented in this appendix. By following the procedure presented in Sec.\ref{SecMethod}, we have evaluated the complex eigenvalues distributions of 10 different realizations of 2000 uniformly random distributed scatterers on a plane. Moreover, the spectral and optical properties of matrix (\ref{Green}), which takes into account the vector nature of light, were compared with those of its scalar approximation \cite{SkipetrovPRL}
\begin{equation}\label{Greenscalar} 
G_{ij}=i\delta_{ij}+(1-\delta_{ij})\frac{e^{ik_0r_{ij}}}{k_0r_{ij}}
\end{equation}

\begin{figure*}[h!]
\includegraphics[width=12cm]{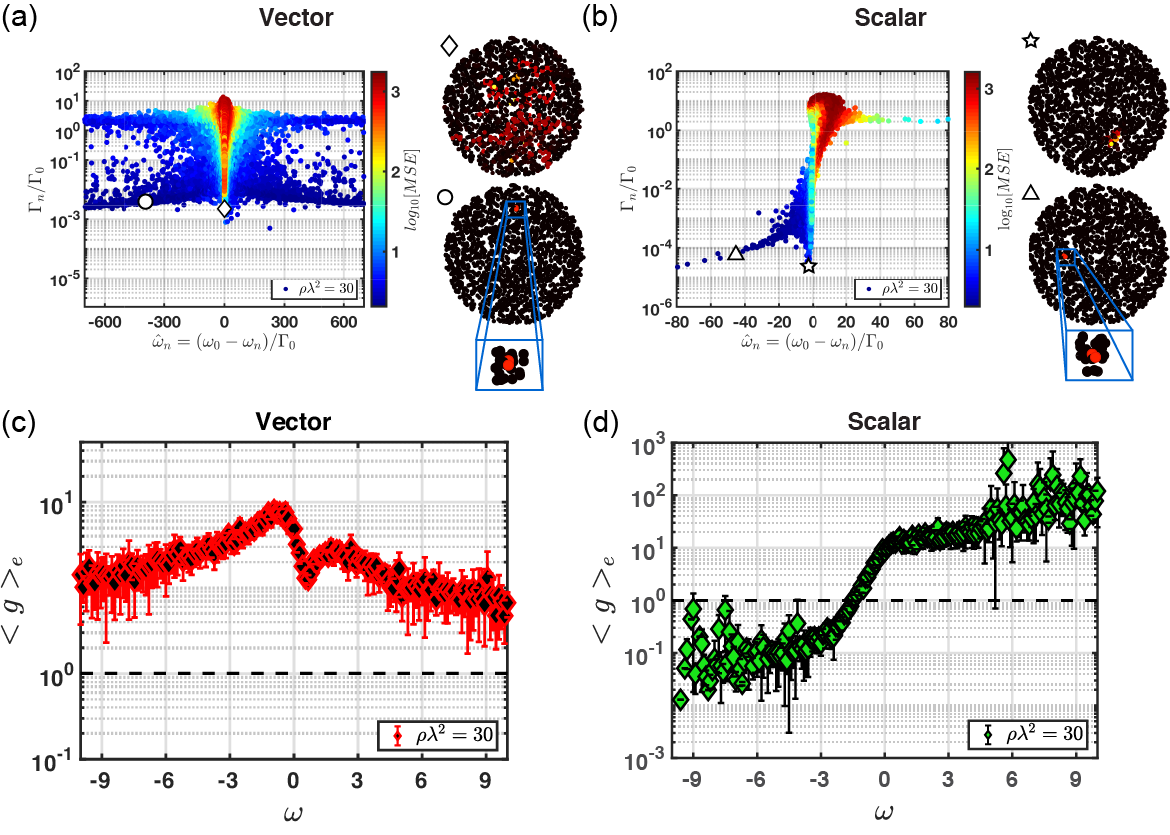}
\caption{Panel (a) and (b) display the complex eigenvalues distributions of 10 different random realizations of the Green's matrix defined by matrix (\ref{Green}) and matrix (\ref{Greenscalar}), respectively. The data are color-coded according to the log10 values of the MSE. Insets: spatial profiles of representative quasi-modes. 
Panel (c) and (d) show the frequency dependence of the Thouless conductance g for the vector and scalar model, respectively. These data are produced by fixing $\rho\lambda^2$=30. The dashed-black line identifies the threshold of the diffusion-localization transition $g$=1. The error bars are calculated as the standard deviation of the Thouless conductance $g$ evaluated for the different disorder realizations.}
\label{FigA}
\end{figure*}

Fig.\ref{FigA} (a-b) display the complex eigenvalues distributions obtained after diagonalizing the matrix (\ref{Green}) and its scalar approximation (\ref{Greenscalar}) for $\rho\lambda^2$=30, respectively. In random media, long-lived resonances do not appear when the vector nature of light is taken into account. Consistently, the spatial profiles of the Green's matrix eigenvectors corresponding to the resonances with the lower decay rates are delocalized across all the structure (see the representative quasi-mode shown in the inset of Fig.\ref{FigA}(a). On the other hand, the situation is completely different in the scalar configuration. Long-lived resonances are clustered around one band of localized quasi-modes near $\hat{\omega}_n\sim$-2.5. The spatial distributions of quasi-modes of the Green's matrix corresponding to this ``tail region" are localized between several particles, as shown in the inset of Fig.\ref{FigA}(b) for a representative scattering resonance (star-marker).

This analysis, inspired by Ref.\cite{SkipetrovPRL}, is confirmed by the frequency dependence of the Thouless conductance $g$. The conductance is evaluated by using eq.(\ref{Thouless}), which has been modified to take into account the effect of the different disorder realizations \cite{SkipetrovPRL,SkipetrovPRB}. Moreover, the contribution of sub-radiant resonances (for which $MSE$=2) is omitted from this analysis \cite{SkipetrovPRL,Bellando,SkipetrovPRB}. As expected, Fig.\ref{FigA} (c) shows that the Thouless conductance $g$ is always larger than one when the vector nature of light is taken into account. On the contrary, the frequency dependence of $g$ shows a transition from $g<$1 to $g>$1 in the scalar case (see (Fig.\ref{FigA}(d))). These data are obtained by fixing $\rho\lambda^2$=30. This analysis confirms the results of Refs.\cite{SkipetrovPRL,Bellando} obtained for a 3D random distribution of electric dipoles. However, in our case localization is less pronounced if compared to the case treated in \cite{SkipetrovPRL,Bellando} for the scalar model. This is due to the fact that in open random 2D arrays leakage through the system plane results in more lossy channels if compared with the corresponding 3D case.
\section{First-neighbor probability density function analysis}\label{Poisson}\label{AppxB}
\begin{figure*}[h!]
\includegraphics[width=12cm]{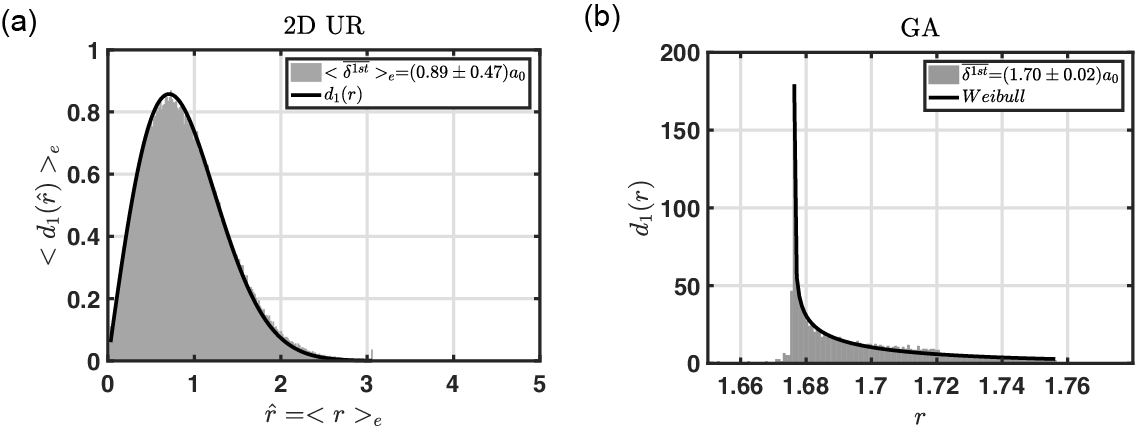}
\caption{Panel (a) and (b) displays the first-neighbor probability density function an homogeneous Poisson process and a GA-spiral, respectively. The black lines are the corresponding analytical density functions obtained by using Eq(\ref{Poisson}) and the Weibull distribution, respectively. In the homogeneous Poisson process two hundred different realizations, with the same density of the GA-spiral,are considered. All these data are in units of the scaling factor $a_0$ (see Eq.\ref{1}). }
\label{FigB}
\end{figure*}

In order to gain more insights on why proximity resonances are absent in Vogel spiral point patterns we study the properties of the first-neighbor probability density function of a GA-spiral as compared to homogeneous Poisson point pattern. It is important to remember that the first-neighbor probability density function is a measure of the spatial uniformity of a given point pattern \cite{DalNegro_Crystals,Illian}. Fig.\ref{FigB} panels (a) and (b) show the results of this analysis as a function of the spacing parameter $r$. Fig.\ref{FigB}(a) is the result of an average over 200 different homogeneous Poisson patterns with exactly the same density of the GA-spiral. The results of Fig.\ref{FigB} clearly demonstrate that the GA-spiral is characterized by a more regular structure as compared to random media. Indeed, the probability density function of  a GA-spiral is extremely peaked around the mean value of the first-neighbor distances and it is very well reproduced by considering a Weibull distribution fitting function, as highlighted by the black-line of Fig.\ref{FigB}(b). On the contrary, the 2D UR configuration is characterized by a Poissonian first-neighbor distribution described by the analytical expression \cite{Illian}
\begin{equation}\label{Poisson}
d_1(r)=\frac{2(\lambda\pi r^2)}{r} e^{-\lambda\pi r^2}
\end{equation}
where $\lambda$ is the intensity of the Poisson point process. It is important to emphasize that the trend of $<d_1(\hat{r})>_e$ ($<\cdots>_e$ indicates the average over an ensemble) in Fig.\ref{FigB}(a) is not the result of a fitting procedure. Rather, it is obtained by using  Eq.(\ref{Poisson}) after calculating $\lambda$ as $N/(\pi\ R^2)$. Here N is the number of points equal to 2000 while $R$ is the maximum radial coordinate of the system (see the insets of Fig.\ref{FigA}). 

Fig.\ref{FigB} clearly shows that two extremely different first-neighbor probability density functions characterize the two considered point processes. 
For a given optical density, the probability of finding two particles very close together is much larger for homogeneous random patterns (see the trend of $<d_1(\hat{r})>_e$ near to $r$=0). On the contrary, proximity resonances do not influence Vogel spirals thanks to these peculiar geometrical properties.
\section{Representative eigenvectors of different localized-resonance bands} \label{AppxC}
\begin{figure*}[h!]
\includegraphics[width=12cm]{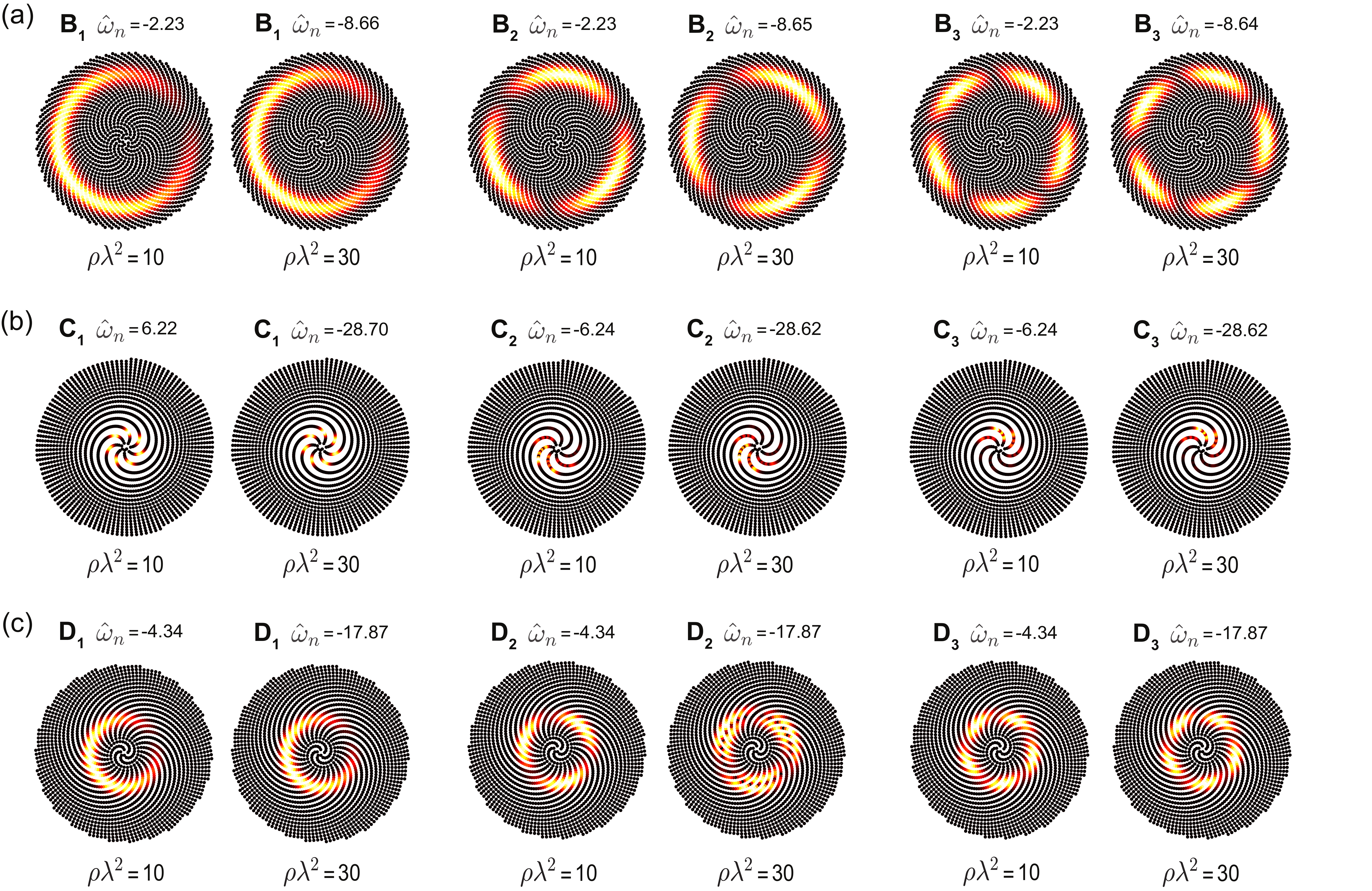}
\caption{Representative spatial distributions of the Green's matrix eigenvectors that belong to the class of scattering resonances that produce the stronger localization feature in the $\tau$-spiral (panel (a)), $\pi$-spiral (panel (b)), and $\mu$-spiral (panel (c)), respectively. B$_j$, C$_j$, and D$_j$ (with j=1,2,3) identify the first three resonances with the lower decay rates ($\Gamma_n/\Gamma_0\ll1$). Moreover, panels (a-c) report these quasi-modes when the optical density $\rho\lambda^2$ is fixed to 10, and 30. The spectral positions of these scattering resonances are identified by the different markers of Fig.\ref{2} (b-d).}
\label{FigC}
\end{figure*}

Figure \ref{FigC} (a-c) display representative eigenvectors corresponding to the different classes of the scattering resonances that lead to more pronounced localization in the $\tau$, $\pi$, and $\mu$ spirals, respectively. The spatial distributions of these quasi-modes correspond to the three eigenvalues of the Green's matrix with the lower decay rates. They are labeled B$_j$, C$_j$, and D$_j$ (with $j$=1,2,3) in the $\tau$, $\pi$, and $\mu$ configurations respectively. In each panel of Fig.\ref{FigC} the optical density is fixed to 10, and 30. Moreover, for each of them, the frequency $\hat{\omega}_n$ is also reported. We clearly observe that exactly the same spatial profile is retrieved when scanning along the dispersion branches identified by the different markers of Fig.\ref{Fig2}. The effect of increasing the optical density $\rho\lambda^2$ is simply to produce a frequency shift in the complex scattering plane, as discussed in the main text for the GA-spiral. Moreover, we notice that the spatial profiles reported in Fig.\ref{FigC} agree very well with what previously reported based on the FEM technique \cite{TrevinoOpticsVogel,Bettotti}. This analysis demonstrate that the different localized resonances of Fig.\ref{Fig2} are the different localized band-edge modes produced by the strongly-fluctuating (multi-fractal) dispersion of the density of states in the different investigated Vogel spirals \cite{TrevinoOpticsVogel,Bettotti}. 

\section{Different coupling terms of the dyadic Green propagator}\label{AppxD}

\begin{figure*}[h!]
\includegraphics[width=12cm]{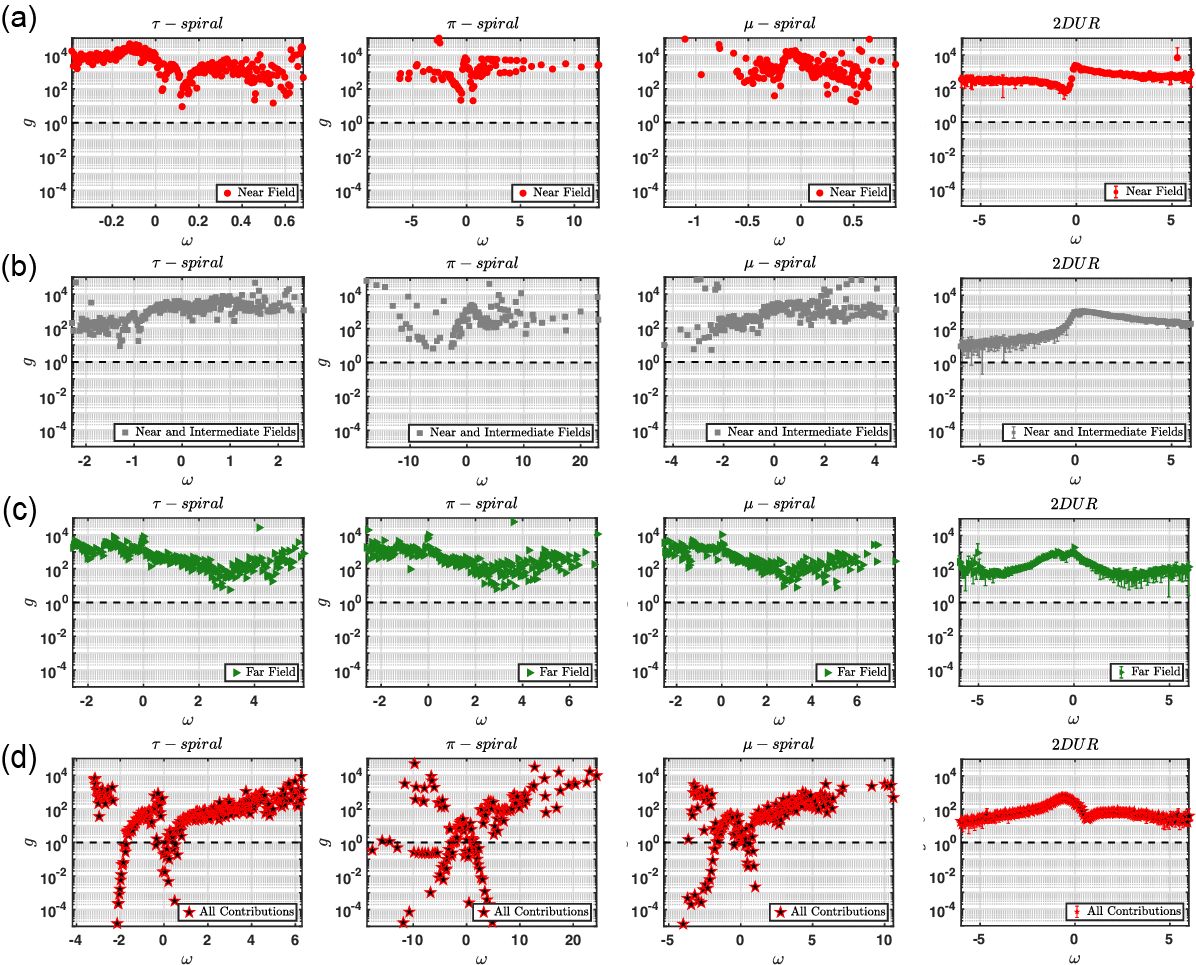}
\caption{Semi-log plots of the Thouless conductance, as a function of $\omega$, obtained by using Eq.(\ref{Thouless}) after diagonalizing the 3$N$$\times$3$N$ Green's matrix associated to only the near-field term (panel (a)), the near-field term plus the intermediate-field regime (panel (b)), the far-field term only (panel (c)), and all the coupling contributions (panel (d)) for the $\tau$, $\pi$, $\mu$ spirals as compared to the 2D UR configuration. This analysis is performed by fixing $\rho\lambda^2$ equal to 10. The error bars are evaluated as the standard deviations of the Thouless conductances calculated for 10 different disorder realizations. The dashed-black lines identify the threshold of the diffusion-localization transition $g$=1.}
\label{FigD}
\end{figure*}

The effects of the different coupling terms of the dyadic Greens's propagator are analyzed for the $\tau$, $\pi$, $\mu$ Vogel spirals as compared to the 2D UR configuration. Fig.\ref{FigD} displays the frequency dependence of the Thouless conductance $g$ obtained by using Eq.(\ref{Thouless}) after diagonalizing the 3$N\times$3$N$ Green's matrix associated to only the near-field term (panel (a)), the near-field plus the intermediate-field contribution (panel (b)), the far-field term only (panel (c)), and all the coupling contributions (panel (d)). These results are obtained for $\rho\lambda^2$=10. Light localization, characterized by $g<$1, only occurs in Vogel spirals when all the coupling terms, including the near-field regime, are taken into account. This shows that light localization in Vogel spiral arrays composed of point-like scatterers results from a collective coupling effect that involves multiple length scales.

On the other hand, homogeneous planar random media do not show any light-localization transitions when the vector nature of light is taken into account confirming the results of Refs.\cite{SkipetrovPRL,Bellando,Maximo}. Indeed, the localization criterium $g(\omega)<$1 is never satisfied in the 2D UR configuration (see the last column of Fig.\ref{FigD}). Interestingly, the Thouless conductance is larger than one also when the only far-field coupling term is taken into account. Hence our findings clearly demonstrate that the absence of any structural correlations is the main responsible that prevents light localization in uniform random arrays when the vector nature of light is taken into account.
\section{Level spacing statistics in Vogel spirals}\label{AppxE}
Level statistics provides important information about electromagnetic propagation in both closed and open scattering systems. 
Indeed, the concept of level repulsion is related to the transport properties supported by eigenmodes because it indicates the degree of spatial overlap between them \cite{Mehta}. Level repulsion can help to discriminate a transition from a delocalized (presence of level repulsion) to a localized wave transport regime (absence of level repulsion). In Refs.\cite{DalNegro_Crystals,Wang} the distribution of level spacings was investigated in different open scattering systems for different scattering strengths. The presence of the level repulsion is characterized by the derivative of the interpolation function, called critical cumulative probability \cite{Zharekeshev}, while the suppression of level repulsion is indicated by fact that the level spacings is described by the Poisson distribution \cite{Haake,DalNegro_Crystals,Wang}.

\begin{figure*}[h!]
\includegraphics[width=12cm]{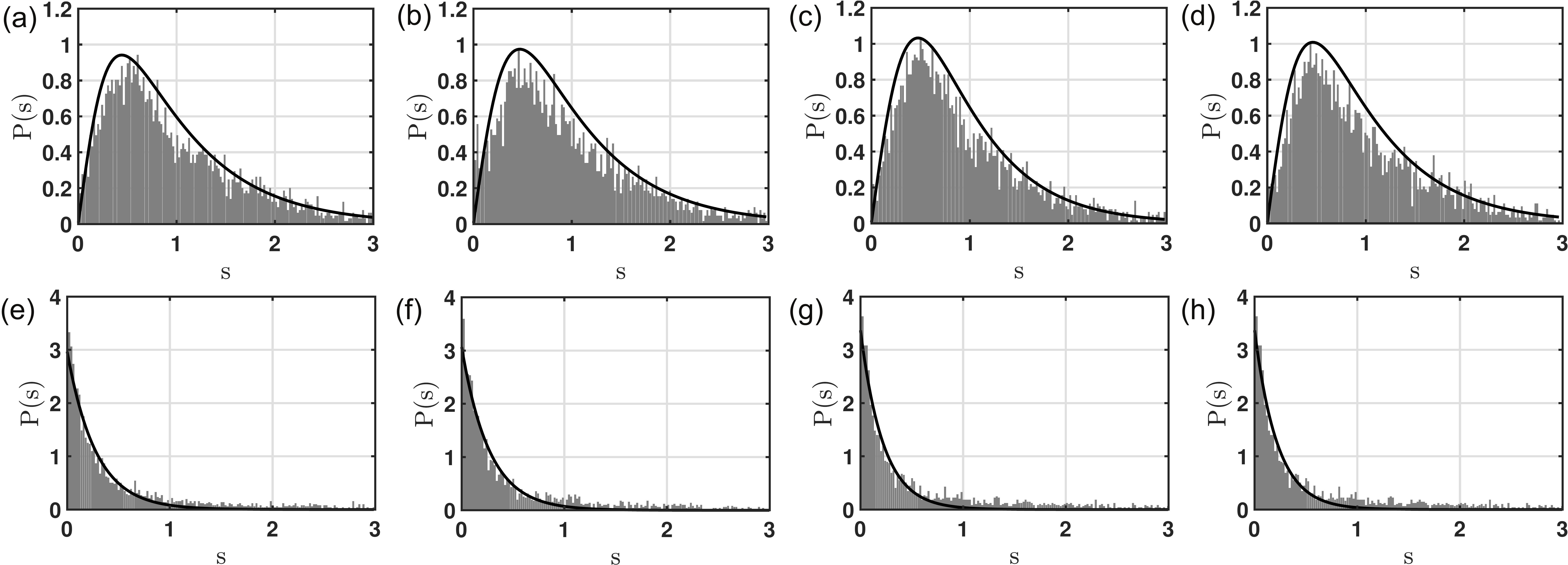}
\caption{Level spacing statistics of the Green's matrix eigenvalues for tow different regime: $\rho\lambda^2=0.1$ (panels a-d) and $\rho\lambda^2=10$ (panels e-h). Panels (a-e), (b-f), (c-g), (d-h) refer to the GA-spiral, $\tau$-spiral, $\pi$-spiral, and $\mu$-spiral configurations, respectively. The fitting curves are performed by using the critical cumulative distribution \cite{Zharekeshev,DalNegro_Crystals,Wang} (solid curves in panels (a-d)) and the Poisson distribution (solid curves in panels (e-h)).}
\label{FigE}
\end{figure*}

The distribution of level spacings is calculated for two different optical density (0.1 Fig.\ref{FigE} (a-d) and 10 Fig.\ref{FigE} (e-f)) for all the investigated Vogel spiral configurations. Fig.\ref{FigE} shows a clear transition between level repulsion at low optical densities and the absence of level repulsion at large optical densities. 
For large optical density, the distribution of level spacings follows the Poisson distribution (no level repulsion), as it occurs for uniform random systems in the localized regime \cite{DalNegro_Crystals}. On the other hand, for weakly scattering systems the level spacing distribution follows the same critical distribution  that it is expected to describe the Anderson transition in random media, where wave functions feature  multi-fractal scaling \cite{Zharekeshev}. Differently from traditional uniform random media where criticality is achieved at the localization threshold, which occurs for a specific optical density in 3D, in Vogel spirals we have verified that the critical behavior occurs for a broader range of optical densities. Interestingly, this behavior was reported also for complex prime arrays \cite{Wang}.

\providecommand{\noopsort}[1]{}\providecommand{\singleletter}[1]{#1}%
\end{document}